\documentclass[12pt,letter,amsmath,amssymb]{article}

\usepackage{graphicx}% Include figure files
\usepackage{color}
\usepackage{dcolumn}% Align table columns on decimal point
\usepackage{bm}% bold math
\usepackage[super,comma,compress]{natbib}

\begin{document}

\title{Raman Spectroscopy of graphene and bilayer under biaxial strain: bubbles and balloons}

\author{Jakob Zabel$^1$, Rahul R. Nair$^2$, Anna Ott$^1$, Thanasis Georgiou$^2$,\\ Andre K. Geim$^2$, Kostya S. Novoselov$^2$, Cinzia Casiraghi$^{*1,3}$}

\maketitle

\paragraph{Affiliation}

$^1$Physics Department, Free University Berlin, Germany\\
$^2$Department of Physics and Astronomy, Manchester University, UK\\
$^3$School of Chemistry and Photon Science Institute, University of Manchester, UK\\

\paragraph{Abstract} In this letter we use graphene bubbles to study the Raman spectrum of graphene under biaxial (e.g. isotropic) strain. Our Gruneisen parameters are in excellent agreement with the theoretical values. Discrepancy in the previously reported values is attributed to the interaction of graphene with the substrate. Bilayer balloons (intentionally pressurized membranes) have been used to avoid the effect of the substrate and to study the dependence of strain on the inter-layer interactions.
\\
\\
Keywords: graphene, bilayer, Raman Spectroscopy, strain, Gruneisen parameters, stacking.

\paragraph{Manuscript text} Graphene is the latest carbon allotrope to be isolated \cite{Nov306(2004),GeimRevNM}, and it is now at
the center of a significant experimental and theoretical research effort~\cite{Nov438(2005),Nov315(2007),Zhang438(2005),CastroNetoRev,geimrev,nobel1,nobel2}. In particular,
near-ballistic transport at room temperature and high carrier mobilities~\cite{MorozovNov(2007),andrei,kimmob,romanNL} make it a potential material for
nanoelectronics.

Strain can be used to tailor the electronic properties of graphene \cite{pereira,guinea}. This allows one to make an all-graphene circuit where all the elements are made of graphene with different amounts and types of strain \cite{pereira}. Furthermore, certain configurations of strain are equivalent to a magnetic field \cite{guinea,guineaprb}, which can be very high \cite{levy}. Thus, it is essential to be able to probe strained graphene and to distinguish between different types of strains.

Elastic and inelastic light scattering are powerful tools for investigating graphene \cite{acfprb,CasiraghiNL2007,GeimAPL}. Raman spectroscopy allows monitoring of doping, defects, disorder, chemical modifications and edges \cite{acfprb,ccapl,Pisana,DasCM,Cedge,Elias,ccPSS,fluoro}. All $sp^2$ bonded carbons show common features in their Raman spectra, the so-called G and D peaks, around 1580 and 1360 cm$^{-1}$~\cite{Ferrari00}. The G peak corresponds to the E$_{2g}$ phonon at the Brillouin zone (BZ) center ($\bf\Gamma$~point). The D peak is due to the breathing modes of six-atom rings and requires a defect for its activation\cite{tk}. It comes from TO phonons around the BZ \textbf{K} point\cite{Ferrari00} and it is activated by an intra-valley scattering process. The 2D peak is the second order of the D peak. This is a single peak in monolayer graphene, whereas it splits into
four bands in bilayer graphene, reflecting the evolution of the band structure \cite{acfprb}. The Raman spectrum of graphene also shows significantly less intensive defect-activated peaks such as the D' peak, which lies at $\sim$ 1620cm$^{-1}$. This is activated by an intra-valley process i.e. connecting two points belonging to the same cone around \textbf{K} (or \textbf{K}')\cite{Cedge}. The second order of the D' peak is called 2D' peak. Since 2D and 2D' peaks originate from a Raman scattering process where momentum conservation is obtained by the participation of two phonons with opposite wavevector (\textbf{q} and -\textbf{q}), they do not require the presence of defects. Thus, they are always visible in the Raman spectrum.

Strain can be very efficiently studied by Raman Spectroscopy since this modifies the crystal phonon frequency, depending on the anharmonicity of the interatomic potentials of the atoms. The rate of this change is given by the Gruneisen parameter ($\gamma$). Several experimental and theoretical works studied uniaxial strain in graphene \cite{huang,acfstrain,yu,ni,hone,janina,strain3}, showing that this leads to: i) softening of the modes for tensile strain; ii) splitting of the G peak for increasing strain; iii) broadening or splitting of the 2D peak, depending on the strain direction. However, there is significant discrepancy among the reported results. The Gruneisen parameters are difficult to study in uniaxial strain because they require the knowledge of the Poisson ratio, which depends on the substrate and on the degree of adhesion between graphene and the substrate itself \cite{acfstrain}. Furthermore, it is difficult to calculate the D and 2D Gruneisen parameters since uniaxial strain moves the relative position of the Dirac cones. This affects the 2D peak phonon because this mode is activated by an inter-valley Raman scattering process \cite{acfstrain}. Finally, the G and 2D Gruneisen parameters strongly depend on the direction of the uniaxial strain and on its strength \cite{hone,janina,strain3}.

Biaxial strain is more suitable for the calculation of the Gruneisen parameter because it does not depend on the Poisson ratio and effects due to the relative anisotropic shifts of the Dirac cones are absent \cite{acfstrain}. The Raman spectrum of graphene under biaxial strain has been experimentally studied only in two works: Ding et al. \cite{ding} used a graphene flake deposited on a piezoelectric substrate. However, here the strain was not directly measured, but derived from the G peak position, using the calculated Gruneisen parameter in Ref. \cite{acfstrain}. The authors also found different Gruneisen parameters for the D and 2D peak, in contrast to the theoretical predictions \cite{acfstrain}. Metzger et al \cite{goldberg} used a suspended graphene over a shallow depression. The strain was directly measured by Atomic Force Microscopy (AFM). In both works the Raman spectrum of graphene does not show any splitting of the G peak and the modes soften for increasing strain. However, very different Gruneisen parameters have been reported, Table 1. We will show that this discrepancy could be due to the strain or doping initially imposed on graphene by the substrate.

In this letter we analyze the Raman spectrum of graphene under biaxial strain by studying graphene bubbles formed during the deposition of large graphene flakes on a oxidized silicon substrate (Si/SiO$_x$). Single layer and bilayer bubbles have been measured and mapped by Raman spectroscopy and AFM. From the shift of the Raman peaks, we derived the Gruneisen parameters for G, D, 2D and 2D' peaks. Furthermore, we also studied the evolution of Raman spectra of bilayer graphene with biaxial strain by pressurizing a graphene membrane with nitrogen gas. Here bilayer is specially chosen to understand the effect of biaxial strain on inter-layer interaction of AB- stacked bilayer graphene. Moreover this technique allowed us to eliminate any effect from strain or doping imposed on graphene by the substrate.

Large single layer graphene flakes (lateral size above 0.1$mm$) were produced by micro-mechanical cleavage of bulk graphite and deposited on Si/SiO$_x$, previously cleaned by oxygen plasma \cite{thanasis}. The bubbles can be easily identified by optical microscope because they are characterized by Newton rings, i.e. fringes of different colors, produced by the interference of the reflected and transmitted light waves between graphene and the silicon \cite{thanasis}. The bubbles have different shape and size \cite{thanasis}: we selected only large and spherical bubbles (average diameter of 5-10 $\mu$m). Graphene and bilayers bubbles have been identified by optical contrast \cite{CasiraghiNL2007,GeimAPL} and by the shape of the 2D peak in the Raman spectrum \cite{acfprb}.

Atomic Force Microscopy has been used to measure the strain ($\epsilon$). %and the pressure in the bubble.
Figure 1 shows an AFM image of a bubble. The strain has been calculated in the following way: from the AFM we know the height (h) and width (w) of the bubble. Assuming that the strain is zero when there is no bubble (h=0), the strain will be: $\epsilon= (L-L_0)/L_0$, where L is the length of the arc after deformation and $L_0$ is the initial length of the arc, i.e. $L_0= w$. We found that for most of the graphene bubbles $\epsilon\sim 1\%$. From the shape of the bubble, it is possible to calculate the differential pressure by using the following relation: $\Delta p= f(\nu)Et[h^3/(w/2)^4]$, where $f(\nu)$ is a function of the Poisson ratio $\nu$, E is the Young's modulus and t is the thickness of the bubble. We used Et=347 N/m and $\nu$=0.16, thus $f(\nu) = 3.09$ \cite{koenig}. We estimated an average pressure of about 15 bar for most of the bubbles.

 %The pressure inside the bubble has been calculated using xxx.

\begin{figure}
\centerline{\includegraphics [width=0.5\textwidth]{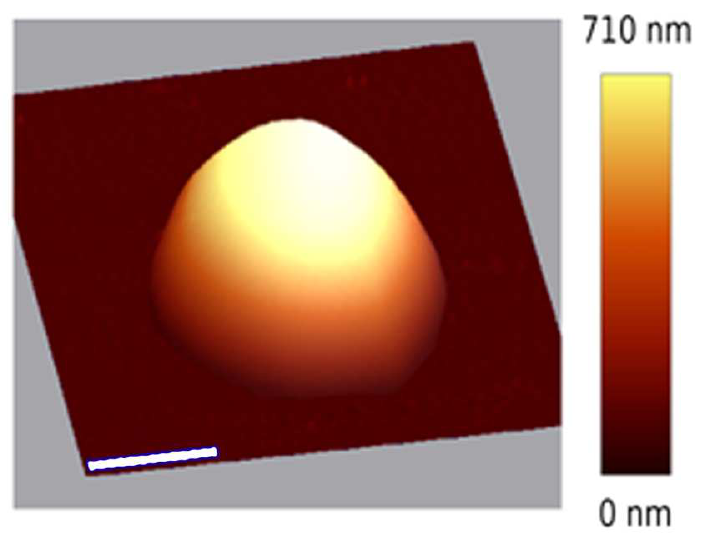}}
\caption{\label{Fig1} (color on-line) AFM image of a single-layer graphene bubble. The lateral scale is 4 $\mu m$.}
\end{figure}

\begin{figure}
\centerline{\includegraphics [width=0.5\textwidth]{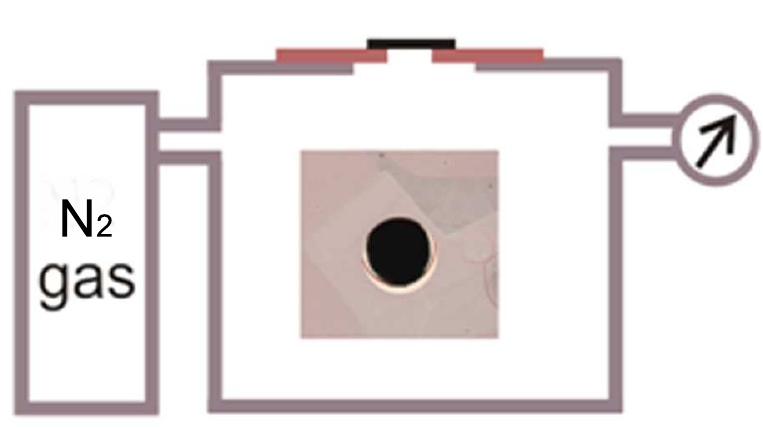}}
\caption{\label{Fig2} (color on-line) Schematic showing the experimental setup used to make the balloons; the inset shows an optical micrograph of a large bilayer covering a 50$\mu m$ aperture in a copper scaffold.}
\end{figure}

A Witec Raman spectrometer, equipped with 488nm and 514.5nm laser lines and a objective with numerical aperture of 0.95, has been used. This gives a laser spot size of about 400nm. The laser power is kept well below 1mW. Careful control of the laser power is necessary: the bubble can move or disappear if the laser power is too high because mass transport under the bubble is activated by laser heating \cite{flynn}. The Raman spectrometer is also equipped with a piezoelectric stage, which we used to Raman map the bubble and to take measurements exactly at the center of the bubble.
The peaks have been fitted with a single Lorenzian lineshape and we analyzed the position and Full Width at Half Maximum (FWHM) of the peaks.

The Gruneisen parameter has been calculated as \cite{acfstrain}: $\gamma= [\omega-\omega_0]/[2\epsilon\omega_0]$, where $\omega$ and $\omega_0$ are the Raman frequencies at finite strain and zero strain, respectively.

Bilayer graphene balloons were produced by pressurizing a specially made metallic container covered with large bilayer graphene membranes with nitrogen gas. Large bilayer graphene membranes were prepared by the method previously reported \cite{booth}. In brief, graphene crystals were exfoliated on top of PMMA coated (90 nm) silicon wafer by mechanical exfoliation. A thick copper film was deposited on top of the selected flakes by using a series of photolithography and electrodeposition steps. The deposited copper film contained an open aperture of 50$\mu m$ in diameter, which was aligned with the chosen graphene crystal so that graphene fully covered the aperture. Copper scaffold with graphene layers were released from PMMA by dissolving in acetone. The samples were finally dried in a critical point dryer to prevent membrane from rupturing because of the surface tension. These graphene membranes were used to seal a small opening in a specially made metallic container with incorporated gas inlet and pressure gauge, Figure 2. Silver epoxy was used to attach the scaffold with graphene to the metallic container. Pressure inside the container was changed controllably by using nitrogen gas. We used a Renishaw Raman spectrometer (514.5nm) to monitor the changes in the Raman spectrum of the balloon under various differential pressures ($dp=P_{inside}-P_{out}$). The changes in Raman spectra of graphene balloons with different pressures were spatially uniform all over the membrane and were completely reversible (see Supporting Information).  We were able to apply a maximum differential pressure of 2 bar. At higher pressures the membranes have a tendency to crack or burst. We speculate that this upper limit of pressure was not limited by intrinsic property of graphene rather due to the weak adhesion of graphene to the metal scaffold or due to the extra strain at the edge of the aperture.

\begin{figure}
\centerline{\includegraphics [width=0.8\textwidth]{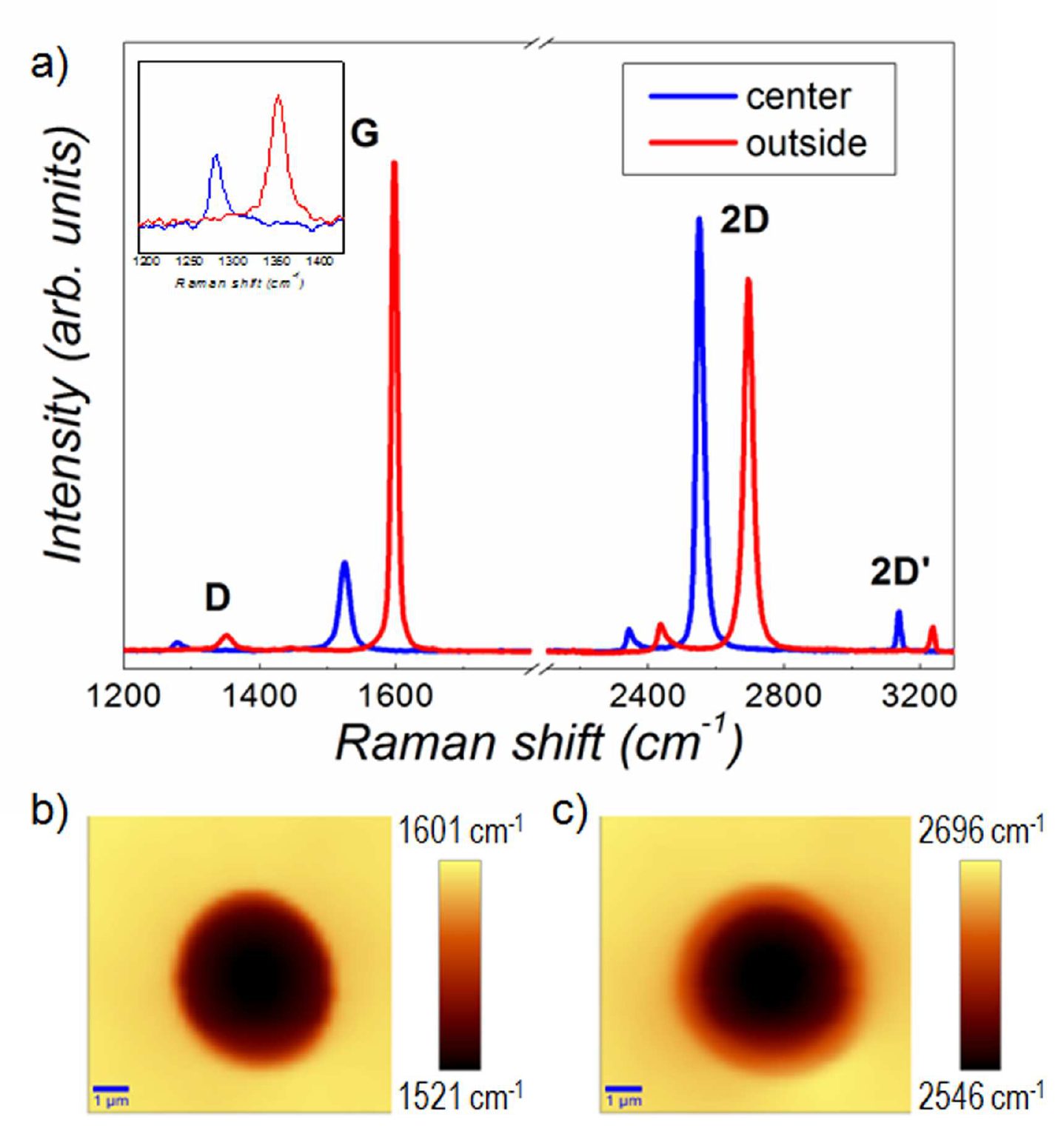}}
\caption{\label{Fig1} (color on-line)(a) Raman spectra measured at the center of a single-layer bubble and on the flat area on the substrate at 488nm;(b) Raman map of the G peak position; (c) Raman map of the 2D peak position. The 2D and G peak positions strongly decrease when moving from the substrate to the center of the bubble.}
\end{figure}

Figure 3(a) shows the Raman spectrum of graphene measured on two adjacent regions: on the substrate, far from the bubble, and at the center of the bubble. A very small D peak is visible in both spectra, as shown in the inset. We can observe that the Raman spectrum taken at the center of the bubble is strongly blue-shifted compared to the Raman spectrum measured on the substrate. No splitting of the G peak and strong broadening/splitting of the 2D peak are observed on the bubble. In particular, the G peak position decreases from 1598 to 1525 cm$^{-1}$, while the 2D position moves from 2695 cm$^{-1}$ to 2552 cm$^{-1}$ when moving from the substrate to the center of the bubble, Figure 3(b,c). The 2D FWHM stays constant at about 30 cm$^{-1}$. These changes in the Raman spectrum are in agreement with biaxial strain \cite{ding}. Note that the shifts in the peak position cannot be explained by doping, as the G peak position for undoped graphene is about 1582 cm$^{-1}$ and it increases for increasing doping level \cite{Pisana}.

\begin{figure}
\centerline{\includegraphics[width=0.55\textwidth] {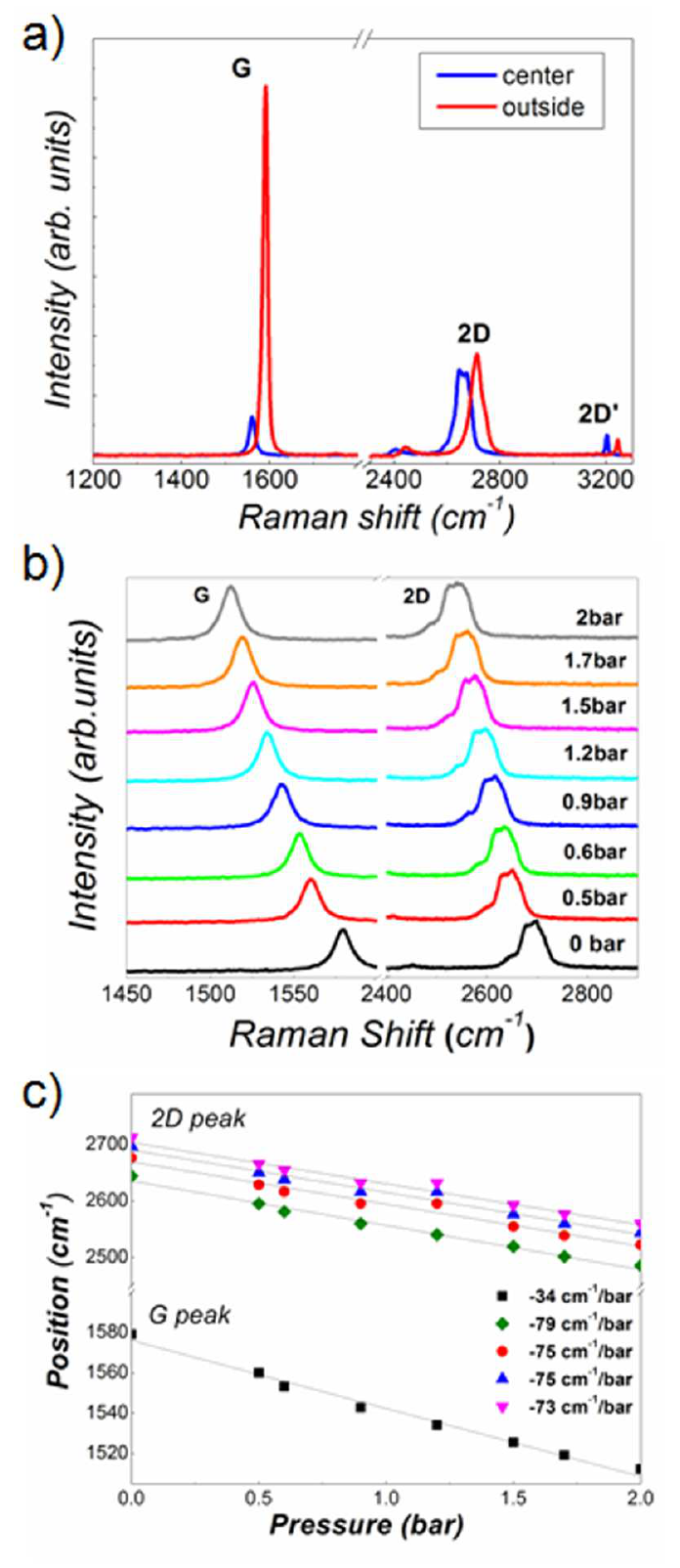}}
\caption{\label{Fig2New} (color on-line) a) Raman spectra measured at the center of the bilayer bubble and on the flat area on the substrate at 488nm; (b) Raman spectra of a bilayer graphene balloon measured for different differential pressure at 514.5nm; (c) G and 2D peak position of a bilayer graphene balloon under different differential pressure.}
\end{figure}

We can now calculate the Gruneisen parameters by using the initial peak frequency measured for unstrained and undoped graphene. We used the following parameters (taken at 488nm): $\omega_0(G)$= 1582cm$^{-1}$; $\omega_0(2D)$= 2692cm$^{-1}$; $\omega_0(2D')$= 3245cm$^{-1}$ and $\omega_0(D)$= 1349cm$^{-1}$ \cite{ccapl,venez}. We then find the following peak shifts: -57cm$^{-1}$/$\%$ for the G peak, -140 cm$^{-1}$/$\%$ for the 2D peak; -108 cm$^{-1}$/$\%$ for the 2D' peak and -68 cm$^{-1}$/$\%$ for the D peak. They give the following Gruneisen parameters: $\gamma(G)$= 1.8; $\gamma(2D)$= 2.6; $\gamma(2D')$= 1.66 and $\gamma(D)$= 2.52, in excellent agreement with previous calculations \cite{acfstrain}, Table 1. In contrast to Ref. \cite{ding} we do not observe any strong difference between $\gamma(2D)$ and $\gamma(D)$ and no splitting of the D peak has been observed, inset Fig. 3(a). On the other side, our $\gamma(2D')$ is in good agreement with the results reported in Ref. \cite{ding}.

In the calculation of the Gruneisen parameters we did not use as initial peak frequencies the values measured on the substrate because it is well know that graphene on Si/SiO$_x$ can be strongly doped \cite{ccapl}. Recent experiments have suggested the presence of a water layer between graphene and the substrate, which could cause strong p-doping \cite{thanasis}. This layer should be absent under the bubble \cite{thanasis}. Indeed, the Raman spectrum measured at the center of the bubble shows a very small G peak intensity, compared to the 2D intensity, typical of undoped graphene \cite{acfprb,ccprb}. In contrast, the Raman spectrum of graphene on the substrate shows: a very high G peak position (1598 cm$^{-1}$) and an intensity ratio between G and 2D peak of about 1. This further confirms that graphene is strongly doped \cite{ccapl,ccprb,cckirc}. By using 1598 cm$^{-1}$ as G peak position for unstrained graphene, we would get a Gruneisen parameter of 2.3, giving an error of 30\%.

Table 1 includes the most recent results on biaxial strain in graphene. We also report the ratio between the phonon peaks shift ($\Delta$). This parameter does not require knowledge of the strain, so it can be used to compare the results of different works.

We now focus on bilayer graphene bubbles. Figure 4(a) shows the Raman spectrum of such an object, measured on two adjacent regions: on the substrate, far from the bubble, and at the center of the bubble. No D peak is visible in both spectra. As seen for the graphene bubble, the Raman spectrum taken at the center of the bilayer graphene bubble is strongly blue-shifted compared to the Raman spectrum measured on the substrate. The G peak moves from 1590 to 1560 cm$^{-1}$. Note the shape of the 2D peak: the Raman spectrum measured on the bubble shows the typical features of an AB-stacked bilayer graphene \cite{acfprb}, in contrast to the Raman spectrum measured on the substrate. Here the 2D peak is almost a broad single-peak and the G peak intensity is very high, compared to the 2D intensity. This clearly shows that the bilayer graphene on the substrate is doped \cite{acf2l}. Effect of strain imposed by the substrate is also possible. Thus, in order to calculate the peak shifts, we used the following values: 1580 cm$^{-1}$ for the G peak and 2664.5, 2696, 2715 and 2728 cm$^{-1}$ for the four 2D components \cite{acf2l}. We then found a shift of - 20cm$^{-1}$ for the G peak and -50cm$^{-1}$ for the 2D peak. This gives a shift ratio between 1.9 and 2.5, depending on the 2D phonon component.

\begin{table}
\centering % centering table
\caption{Table reporting the main results obtained for biaxial strain in graphene. Note that Ref. \cite{ding} do not measure directly the Gruneisen parameter of the G peak, but use the value reported in Ref.\cite{acfstrain}. The error bars take into account the uncertainty in the peak positions. $\Delta(2D/G)$ is the ratio between the shift of the 2D and G peaks.}

\begin{tabular}{l ccccc}
\hline
& $\Delta(2D/G)$ & $\gamma$(G)&$\gamma$(2D)&$\gamma$(D)&$\gamma$(2D') \\
\hline\hline
from uniaxial \cite{acfstrain} & 3.03 &1.99 &3.55 & NA&NA\\
Calculated \cite{acfstrain} &2.48&1.8&2.7&2.7&NA\\
Calculated \cite{strain3} &NA&1.89&NA&NA&NA\\
Calculated \cite{janina} &2.25&NA&NA&NA&NA\\
Depression \cite{goldberg}&2.63&2.4&3.8&NA&NA\\
Piezo \cite{ding} &2.8&[1.8]&2.98&2.3&1.73\\
Graphene bubble &2.45$\pm$15\%&1.8$\pm$10\%&2.6$\pm$5\%&2.52$\pm$9\%&1.66$\pm$5\%\\
Bilayer bubble &1.9-2.5&&&&\\
Bilayer balloon &2.15-2.32&&&&\\
\hline\hline
\end{tabular}
\label{tab:hresult}
\end{table}

In order to avoid any effect on the Raman spectrum from doping or strain imposed by the coupling with the substrate, we investigated the Raman spectrum of bilayer graphene balloons. Figure 4(b) shows the first and second order Raman spectra measured at different pressure. Figure 4(b,c) shows that the Raman spectrum rigidly blue-shifts for increasing pressure; furthermore, the 2D peak shape does not change with the pressure. We found that the shift in the G peak and 2D peak is 34 cm$^{-1}$/bar and 73-79 cm$^{-1}$/bar for the 2D components, Figure 4(c). This gives a peak shift ratio between 2.15 and 2.32, in good agreement with the results of the bilayer graphene bubble. Assuming a Gruneisen parameter of 1.8 for the G peak, we find that the maximum strain reached in the balloon is about 1.2\%, which is too small to break the balloon or to strongly alter the electronic structure of bilayer graphene. This value is also in good agreement from the strain calculated by using $w/2= 25 \mu m$ and h=3.3 $\mu m$, where h is obtained from the formula linking the differential pressure with the bubble shape (here Et= 694 N/m being a bilayer \cite{koenig}).

In conclusion, we have used graphene bubbles to study the Raman spectrum of graphene under biaxial strain. We have calculated the Gruneisen parameters of the G, 2D, D and 2D' peaks: they are in excellent agreement with previously calculated parameters. We have also investigated the Raman spectrum of bilayer graphene bubbles and balloon: we found that in both cases the strain does not alter the AB-stacking configuration, probably because the strain is too small ($\sim$ 1.2\%). We have shown that initial doping or strain caused by the interaction of graphene with the substrate can strongly affect the measured Gruneisen parameters. This could explain the discrepancy in the previously reported Gruneisen parameters.

Acknowledgement: this work is partially funded by the Alexander von Humboldt Foundation in the framework of the Sofja Kovalevskaja Award, endowed by the Federal Ministry for Education and Research of Germany.

Supporting Information Available: Raman spectra taken on multiple spots of a pressurized graphene balloon; Raman spectra of a depressurized membrane, and some micrographs of a bursted bilayer balloon are available free of charge via the Internet at http://pubs.acs.org.


\begin{thebibliography}{99}

\bibitem{Nov306(2004)}K. S. Novoselov, A. K. Geim, S. V. Morozov, D. Jiang, Y. Zhang, S. V.Dubonos, I. V. Grigorieva, A. A. Firsov; Science, \textbf{306,} 666 (2004).

\bibitem{GeimRevNM}A. K. Geim, K. S. Novoselov; Nature Materials, \textbf{6,} 183 (2007).

\bibitem{Nov438(2005)}K. S. Novoselov, A. K. Geim, S. V. Morozov, D. Jiang, M. I. Katsnelson, I. V.
    Grigorieva, S. V. Dubonos, and A. A. Firsov; Nature (London), \textbf{438,} 197 (2005).

\bibitem{Nov315(2007)}K. S. Novoselov, Z. Jiang, Y. Zhang, S. V. Morozov, H. L. Stormer, U.
    Zeitler, J. C. Maan, G. S. Boebinger, P. Kim, and A. K. Geim; Science, \textbf{315,} 1379 (2007).

\bibitem{Zhang438(2005)} Y. Zhang, Y.W. Tan, H. L. Stormer, and P. Kim; Nature (London), \textbf{438,} 201 (2005).

\bibitem{CastroNetoRev}A. H. Castro Neto, F. Guinea, N. M. R. Peres, K. S. Novoselov, A. K. Geim; Rev. Mod. Phys. \textbf{81}, 109 (2009).

\bibitem{geimrev} A. K. Geim, Science, \textbf{324}, 1530 (2009).

\bibitem{nobel1} K. S. Novoselov, Reviews of Modern Physics \textbf{83}, 837 (2011).

\bibitem{nobel2} A. K. Geim, Reviews of Modern Physics \textbf{83}, 851 (2011).

\bibitem{MorozovNov(2007)}S. V. Morozov, K. S. Novoselov, M. I. Katsnelson, F. Schedin, D. C. Elias, J. A. Jaszczak, A. K. Geim; Phys. Rev. Lett., \textbf{100,} 016602 (2008).

\bibitem{andrei} X. Du, I. Skachko, A. Barker, E. Y. Andrei, Nature Nanotech., \textbf{3}, 491 (2008)

\bibitem{kimmob}  K. I. Bolotin, K. J. Sikes, J. Hone, H. L. Stormer, P. Kim  arXiv:0802.2389; K. I. Bolotin, K. J. Sikes, J. Hone, H. L. Stormer, P. Kim, Phys. Rev. Lett., 101, 096802 (2008).

%\bibitem{Han}M. Y. Han, B. Özylmaz, Y. Zhang, P. Kim; Phys. Rev. Lett., \textbf{98,} 206805 (2007).
%10
%\bibitem{Chen}Z. Chen, Y.M. Lin, M. Rooks, P. Avouris; Physica E, \textbf{40,} 228 (2007).

%\bibitem{Zhang86}Y. Zhang, J. P. Small, W. V. Pontius, P. Kim; Appl. Phys. Lett., \textbf{86,} 073104 (2005).

%\bibitem{Lemme}M. C. Lemme, T. J. Echtermeyer, M. Baus, and H. Kurz; IEEE El. Dev. Lett.,\textbf{28,} 4 (2007).

\bibitem{romanNL} A. S. Mayorov, R. V. Gorbachev, et al., Nano Lett. \textbf{11}, 2396 (2011).

\bibitem{pereira} V. M. Pereira, A. H. Castro Neto, Phys. Rev. Lett. \textbf{103}, 046801 (2009)

\bibitem{guinea} F. Guinea, M. I. Katsnelson, A. K. Geim, Nature Phys. \textbf{6}, 30 (2010)

\bibitem{guineaprb} F. Guinea, A. K. Geim, et al., Phys. Rev. B \textbf{81}, 035408 (2010).

\bibitem{levy} N. Levy, S. A. Burke, K. L. Meaker, M. Panlasigui, A. Zettl, F. Guinea, A. H. castro Neto, M. F. Commie, Science, \textbf{329}, 544 (2010)

\bibitem{acfprb}A. C. Ferrari, J. C. Meyer, V. Scardaci, C. Casiraghi, M. Lazzeri, F. Mauri, S. Piscanec, Da Jiang, K. S. Novoselov, S. Roth, A. K. Geim; Phys. Rev. Lett., 97, 187401 (2006).

\bibitem{CasiraghiNL2007}C. Casiraghi, A. Hartschuh, E. Lidorikis, H. Qian, H. Harutyunyan, T. Gokus, K. S. Novoselov, A. C. Ferrari; Nano. Lett., 7, 2711 (2007).

\bibitem{GeimAPL}P. Blake, E. W. Hill, A. H. Castro Neto, K. S. Novoselov, D. Jiang, R. Yang, T. J. Booth, A. K. Geim; Appl. Phys. Lett., 91, 063124 (2007).

\bibitem{ccapl} C. Casiraghi, S. Pisana, K. S. Novoselov, A. K. Geim, A. C.Ferrari; Appl. Phys. Lett., 91, 233108 (2007).

\bibitem{Pisana}S. Pisana, M. Lazzeri, C. Casiraghi, K. S. Novoselov, A. K. Geim, A. C. Ferrari, F. Mauri, Nat. Mat. 6, 198 (2007).

%\bibitem{ACFRamanSSC} A. C. Ferrari; Solid State Comm., 143, 47 (2007).

\bibitem{DasCM} A. Das, S. Pisana, S. Piscanec, B. Chakraborty, S. K. Saha, U. V. Waghmare, R. Yang, H. R. Krishnamurhthy, A. K. Geim, A. C. Ferrari, A. K. Sood Nature Nano. 3, 210 (2008)

\bibitem{Cedge} C. Casiraghi, A. Hartschuh, H. Qian, S. Piscanec, C. Georgi, A. Fasoli, K. S. Novoselov, D. M. Basko, A. C. Ferrari, Nano Lett. 9, 1433 (2009)

\bibitem{Elias} D. C. Elias, R. R. Nair, T. M. G. Mohiuddin, S. V. Morozov, P. Blake, M. P. Halsall, A. C. Ferrari, D. W. Boukhvalov, M. I. Katsnelson, A. K. Geim, K. S. Novoselov, Science 323, 610 (2009).

\bibitem{ccPSS} C. Casiraghi, Phys. Status Solidi- Rapid Research Lett., \textbf{3}, 175 (2009).

\bibitem{fluoro} R. R. Nair, W. Ren, R. Jalil, I. Riaz, V. G. Kravets, L. Britnell, P. Blake, F. Schedin, A. S. Mayorov, S. Yuan \emph{et al}. Small, \textbf{6}, 2877 (2010)

\bibitem{Ferrari00} A.C. Ferrari, J. Robertson Phys. Rev. B \textbf{61}, 14095 (2000); {\it ibid.} \textbf{64}, 075414 (2001).

\bibitem{tk} F. Tuinstra, J. L. Koenig, J. Chem. Phys. \textbf{53}, 1126 (1970)

\bibitem{huang} M. Huang, H. Yan, C. Chen, D. Song, T. F. Heinz, J. Hone, PNAS, \textbf{106}, 7304 (2009)

\bibitem{acfstrain} T. M. G. Mohiuddin, A. Lombardo, R. R. Nair, A. Bonetti, G. Savini, R. Jalil, N. Bonini, D. M. Basko, C. Galiotis, N. Marzari, K. S. Novoselov, A. K. Geim, A. C. Ferrari, Phys Rev. B 79, 205433 (2009)

\bibitem{yu} T. Yu, Z. Ni, C. Du, Y. You, Y. Wang, Z. Shen, Phys. Chem. Lett. C, \textbf{112}, 12602 (2008)

\bibitem{ni} Z. H. Ni, T. Yu, Y. H. Lu, Y. Y. Wang, Y. P. Feng, Z. X. Shen, ACS Nano, \textbf{2},2301 (2008)

\bibitem{hone} M. Huang, H. Yan, T. F. Heinz, J. Hone, Nano Lett. \textbf{10}, 4074 (2010)

\bibitem{janina} M. Mohr, J. Maultzsch, C. Thomsen, Phys. Rev. B \textbf{82}, 201409 (2010)

\bibitem{strain3} Y. C. Cheng, Z. Y. Zhu, G. S. Huang, and U. Schwingenschlögl, Phys. Rev. B \textbf{83}, 115449 (2011)

\bibitem{ding} F. Ding, H. Ji, Y. Chen, A. Herklotz, K. Dorr, Y. Mei, A. Rastelli, O. G. Schmidt, Nano Lett. \textbf{10}, 3453 (2010)

\bibitem{goldberg} C. Metzger, S. Remi, M. Liu, S. V. Kusminskiy, A. H. Castro Neto, Anna K. Swan, B. B. Goldberg, Nano Lett. \textbf{10}, 6 (2010)

\bibitem{thanasis} T. Georgiou, L. Britnell, P.  Blake, R. Gorbachev, A. Gholinia, A. K. Geim, C. Casiraghi, K. S. Novoselov, Appl. Phys. Lett. \textbf{99}, 093103 (2011)

\bibitem{koenig} S. P. Koening, N. G. Boddeti, M. L. Dunn, J. S. Bunch, Nature Nanotech. 6, 543 (2011)

\bibitem{flynn} E. Stolyarova, D. Stolyarov, K. Bolotin, S. Ryu, L. Liu, K. T. Rim, M. Klima, M. Hybertsen, I. Pogorelsky, I. Pavlishin et al, Nano Lett. \textbf{9}, 332 (2009)

\bibitem{booth} T. J. Booth, P. Blake; R. R. Nair, D. Jiang, E. W. Hill, U. Bangert, A. Bleloch, M. Gass, K. S. Novoselov, M. I. Katsnelson, A. K. Geim, Nano Lett., \textbf{8}, 2442 (2008).

\bibitem{venez} P. Venezuela, M. Lazzeri, F. Mauri, Phys. Rev. B \textbf{84}, 035433 (2011)

\bibitem{ccprb} C. Casiraghi, Phys. Rev. B \textbf{80}, 233407 (2009)

\bibitem{cckirc} C. Casiraghi, Phys. Status Solidi B, \textbf{248}, 2593 (2011)

\bibitem{acf2l} W. Zhao, P. H. Tan, J. Liu, A. C. Ferrari, J. Am. Chem. Soc. \textbf{133}, 5941 (2011).



\end{thebibliography}
\end{document}